%% file: main.tex
\documentclass[a4paper,11pt]{article}
\pdfoutput=1
\usepackage{jheppub}

\usepackage{slashed}
\usepackage{amsmath}
\title{{\boldmath 
Two-loop QCD corrections to $ H \rightarrow b + \bar{b} + g $ at higher powers in the dimensional regulator}}

\author[a,b]{Pulak Banerjee,}
\author[c]{Chinmoy Dey,}
\author[d]{M. C. Kumar, }
\author[d]{Vaibhav Pandey}
\author[e]{and V. Ravindran}
\affiliation[a]{School of Physical Sciences, National Institute of Science Education and Research, Jatni, 752050, India}
\affiliation [b]{Homi Bhabha National Institute, Training School Complex, Anushakti Nagar, Mumbai, 400094, India}
\affiliation[c]{Theoretical Physics Division, Physical Research Laboratory, Navrangpura, Ahmedabad 380009, India}
\affiliation[d]{Department of Physics,
	Indian Institute of Technology Guwahati, Guwahati-781039, India}
\affiliation[e]{The Institute of Mathematical Sciences, Taramani, Chennai 600113, India}

\emailAdd{pulak.banerjee@niser.ac.in}
\emailAdd{d.chinmoy@iitg.ac.in}
\emailAdd{mckumar@iitg.ac.in}
\emailAdd{vphiitg@iitg.ac.in }
\emailAdd{ravindra@imsc.res.in}

\abstract{
We compute the two-loop massless QCD corrections to the amplitude of Higgs boson decay to bottom quark pair and a gluon ($H \rightarrow b + \bar{b} + g$) in the higher powers of the dimensional regularization parameter $\epsilon$. The calculation is performed by projecting the amplitude onto the appropriate Lorentz structures related to the process. 
We also show the numerical behaviour of the form factors for a few sample phase-space points.
These amplitudes are necessary ingredients for computing the three-loop virtual corrections to bottom-quark annihilation to Higgs plus jet production at the hadron collisions. 
}

\begin{document}




\maketitle



\section{Introduction}
\label{sec:intro}
The Standard Model (SM) has been tested over decades by several experiments with very high precision. 
The discovery of the Higgs boson by the ATLAS~\cite{ATLAS:2012yve} and CMS~\cite{CMS:2012qbp} collaborations at the Large Hadron Collider (LHC) is an important milestone in establishing the electroweak symmetry breaking mechanism in the SM. 
The generation of the masses of the elementary particles in the SM is explained by the Higgs mechanism ~\cite{Higgs:1964ia,PhysRevLett.13.508,Higgs:1966ev,Englert:1964et,Guralnik:1964eu}.
Studying the CP properties of this discovered boson is important to understand whether or not the discovered particle is the SM Higgs boson~\cite{CMS:2014nkk,ATLAS:2013xga,ATLAS:2020rej,CMS:2020cga,CMS:2021sdq}. 
The mass of the Higgs boson is measured with high precision and is found to be $m_H = 125.04 \pm 0.12$ GeV~\cite{CMS:2024eka}, which is an important parameter in the study of several processes in the SM.

The inclusive Higgs production is the dominant channel at the LHC, which, when measured in the experiment, receives contributions from both the gluon fusion $gg \to H$~\cite{Dawson:1990zj} and the bottom annihilation channels $b\bar{b} \to H$.
The other subdominant channels are vector boson fusion (VBF)~\cite{Rainwater:1997dg}, associated Higgs production with vector boson (VH)~\cite{Brein:2012ne}, through bottom annihilation, and top associated production channels ($t\bar{t}H$)~\cite{Catani:2021cbl,Devoto:2024nhl,Agarwal:2024jyq,Buonocore:2025fqs}.
From the theory side, the Higgs production cross section 
in these channels is known to a very good accuracy with the help of predictions from the perturbation theory in QCD. 
The results up to next-to-next-to-next-leading order (N$^3$LO) in QCD are already available for the gluon fusion channel
~\cite{Dawson:1990zj, Djouadi:1991tka, Graudenz:1992pv, Spira:1995rr, Harlander:2001is, Anastasiou:2002yz, Harlander:2002wh, Catani:2003zt, Ravindran:2003um, Anastasiou:2004xq, Ravindran:2004mb, Harlander:2005rq, Anastasiou:2005qj, Aglietti:2006tp, Anastasiou:2006hc, Ravindran:2006bu, Bonciani:2007ex, Anastasiou:2009kn, Anastasiou:2015vya,Anastasiou:2016cez} , vector boson fusion channel~\cite{Dreyer:2016oyx} and Higgs production in association with a vector boson~\cite{Baglio:2022wzu,Das:2022zie}.  
In the context of the bottom annihilation process $b\bar{b} \to H$, NNLO calculation has been done in~\cite{Harlander:2003ai} and partial three-loop N$^3$LOsv results are available~\cite{Ahmed:2014cha}. 
The Higgs production through bottom annihilation is known till $N^3$LO accuracy~\cite{Duhr:2019kwi}.

Apart from inclusive Higgs boson production cross sections, the measurements of differential distributions such as transverse momentum and rapidity distribution are useful in probing the Higgs couplings to the SM fields.
These differential distributions allow a direct comparison of theoretical predictions with experimental measurements.
An important process in this direction is 
the Higgs production in association with a jet (Higgs $+1$jet) in hadron  collisions~\cite{Chen:2014gva,Boughezal:2015dra,Boughezal:2015aha,Chen:2016zka,Chen:2018pzu,Chen:2019wxf,Campbell:2019gmd}.
%
Higgs $+ 1$ jet production is a key process for experimental analyses involving boosted Higgs bosons and jet-veto observables~\cite{ATLAS:2019nkf,CMS:2025fwn,Banfi:2012vf}.
%
The dominant contribution to the Higgs$+1$jet production in hadron collisions comes from the gluon fusion channel. 
The NNLO QCD calculation for the Higgs$+ 1$jet process in the EFT framework has been done in~\cite{Boughezal:2013uia}.
The next dominant contribution to Higgs$ +1$jet production comes from the bottom annihilation process which is about $3\%$ of that in gluon-fusion channel~\cite{Boughezal:2013uia,Mondini:2021nck}. 
The NLO QCD correction to Higgs $+1$jet production via the bottom annihilation process is available in the literature~\cite{Harlander:2010cz}.
The NNLO QCD correction to this process was computed in~\cite{Mondini:2021nck}.
The observables with jet vetoes are especially useful, as they significantly improve the signal sensitivity and help in studying the properties of the Higgs boson and its interactions with other SM particles \cite{ATLAS:2014yga, CMS:2015qgt}.

Currently, a large amount of data is already available from Run 3 of the LHC, and an even larger dataset is expected from the high-luminosity phase of the LHC.
This requires precision calculations (at least $\mathrm{N^3LO}$ accuracy) to carry out phenomenological studies with about 1\% precision \cite{Caola:2022ayt}. 
One of the important steps in extending the current results for Higgs+jet to $\mathrm{N^3LO}$ is to compute two-loop results to higher powers of the dimensional regulator $\epsilon$.  
These terms contribute to higher-order QCD results and are important ingredients for the three-loop results. 
It is to be noted that for the Higgs production through bottom annihilation, the dependence on the unphysical scales is seen to reduce at N$^3$LO level~\cite{Duhr:2019kwi}.

At the one-loop level, scattering amplitudes exhibit divergences that are typically treated using dimensional regularization. 
In this approach, the calculation is performed in $d = 4 + \epsilon$ dimensions, and the loop amplitudes are written as Laurent expansions in $\epsilon$. 
To achieve the N$^3$LO results requires the knowledge of one-loop amplitudes up to $\mathcal{O}(\epsilon^{4})$, the two-loop amplitudes up to $\mathcal{O}(\epsilon^{2})$, and the three-loop amplitudes up to $\mathcal{O}(\epsilon^{0})$. 
This work provides the first of these key components: the two-loop amplitudes of $H\to b\bar{b}g$, evaluated up to $\mathcal{O}(\epsilon^{2})$ and one loop results $\mathcal{O}(\epsilon^{4})$. 
Previous studies obtained these two loop amplitudes only up to $\mathcal{O}(\epsilon^{0})$~\cite{Ahmed:2014pka,Mondini:2019vub}, which has already played a crucial role in NNLO QCD analyses of Higgs+jet production~\cite{Mondini:2021nck} and in determining the Higgs transverse-momentum spectrum~\cite{Mondini:2021nck,Cal:2023mib} through bottom quark annihilation.

Predictions for Higgs production in association with one jet through an effective gluon--gluon--Higgs interaction in the infinite top-quark mass limit are available at the two-loop level in QCD to higher powers of $\epsilon$ in Ref.~\cite{Gehrmann:2023etk}. 
As experimental data become more precise, it becomes necessary to include subdominant channels, such as bottom-quark annihilation to Higgs+jet. Pseudo-scalar Higgs boson case has already been studied for decay to three partons at two-loop to higher power in $\epsilon$ up to $\mathcal{O}(\epsilon^{2})$ in Ref.~\cite{Banerjee:2017faz,Banerjee:2024zho}. In this work, we present the one and two-loop amplitudes for the process $H \to b \bar{b} g$ to higher powers of $\epsilon$, which can be analytically continued to obtain the amplitudes for 
$b \bar{b} \to H g$, $b g \to H b$ and $\bar{b} g \to H \bar{b}$. 
In this process, the Higgs boson couples to the bottom quarks via the Yukawa coupling $\lambda$ where, the exact bottom quark mass is retained, while neglecting the bottom quark mass in the loops. 



The article is organised as follows. 
In Section 2, we describe the process under consideration and the tensor decomposition of the amplitude.
In Section 3, we describe the calculation and renormalization procedure. 
Finally, we conclude in Section 4.

\section{Higgs decay amplitudes}
  We consider the production of a bottom quark pair in association with a gluon via the decay of a Higgs boson, and for the calculation purpose, we follow the same notation as in Ref.~\cite{Ahmed:2014pka}: 
  \begin{align}
    H(q) \rightarrow b(p_1) + \bar{b}(p_2) + g(p_3),
  \end{align}
  where $q^2 = M_H^2$ and $p_i^2 = 0$ for $i=1,2,3$.
  With Mandelstam variables defined as
  \begin{align}\label{eq:stu}
    s = (p_1 + p_2)^2, \quad t = (p_2 + p_3)^2, \quad u = (p_1 + p_3)^2,
  \end{align}
  We have the relation
  \begin{align}
    s + t + u = M_H^2.
  \end{align}
  We define the scattering amplitudes for this process as
  \begin{align}\label{eq:amplitude}
    \mathcal{A}(p_{1},p_{2},p_{3}) = \mathcal{A}_{\mu}(p_{1},p_{2},p_{3}) \epsilon^{*~\mu }(p_{3}),
  \end{align}
  where $\epsilon^{\mu}$ is the polarization vector of the outgoing gluon.
  The gluon polarization sum is given by
  \begin{align}\label{eq:gluon_polarization_sum}
    \sum_{\lambda} \epsilon_{\lambda}^{*~\mu}(p_{3}) \epsilon_{\lambda}^{\nu}(p_{3}) = -g^{\mu \nu} + \frac{p_{3}^{\mu} n^{\nu} + p_{3}^{\nu} n^{\mu}}{p_{3} \cdot n},
  \end{align}
  where $n^{\mu}$ is an arbitrary light-like four vector. For our calculation we choose $n^{\mu} = p_{1}^{\mu}$.
  
  To calculate the partonic current, it is convenient to express it in a tensor decomposition. 
  Following the tensorial structure of the $\mathcal{A}^{\mu}$, Lorentz invariance and transversality condition ($p_{3} \cdot \epsilon = 0$) of external gluon allows us to write the amplitude as a linear combination of the following set of three tensor structures:
  \begin{align}
    \mathcal{A}^{\mu}(p_{1},p_{2},p_{3}) = \overline{u}(p_{1}) \left[ \mathcal{F}_{1} ~p_{1}^{\mu} + \mathcal{F}_{2}~ p_{2}^{\mu}  + \mathcal{F}_{3} ~\slashed{p}_{3} \gamma^{\mu} \right] v(p_{2}).
  \end{align}
  Further QCD ward identity $\left( p_{3~\mu}\mathcal{A}^{\mu}=0 \right)$ reduces the three form factors ($\mathcal{F}_{1}$, $\mathcal{F}_{2}$ and $\mathcal{F}_{3}$) to only two independent form factors ($A_{1}$ and $A_{2}$):
  \begin{align}\label{eq:tensor_decomposition}
    & \mathcal{A}^{\mu}(p_{1},p_{2},p_{3}) = \overline{u}(p_{1}) \left[ A_{1} \left( p_{2}\cdot p_{3} ~ p_{1}^{\mu} - p_{1}\cdot p_{3} ~ p_{2}^{\mu} \right) + A_{2} ~\slashed{p}_{3} \gamma^{\mu} \right] v(p_{2}).
  \end{align}
  The equation \ref{eq:amplitude} can be written as,
  \begin{align}
    & \mathcal{A}(p_{1},p_{2},p_{3}) \equiv A_{1} T_{1} + A_{2} T_{2}
  \end{align}
  where, 
  \begin{align}
    & T_{1} = \overline{u}(p_{1}) \left[ p_{2}\cdot p_{3} ~ p_{1}^{\mu} - p_{1}\cdot p_{3} ~ p_{2}^{\mu} \right] v(p_{2}) \epsilon^{*}_{\mu}(p_{3}),
    & T_{2} = \overline{u}(p_{1}) \slashed{p}_{3} \gamma^{\mu} v(p_{2}) \epsilon^{*}_{\mu}(p_{3}).
  \end{align}
  The original form-factors $\mathcal{F}_{i}$ can be written in terms of $A_{i}$ as, $\mathcal{F}_{1} = (p_{2}\cdot p_{3})~ A_{1}$ and $\mathcal{F}_{3} = A_{2}$.
  We note that the tensor decomposition (\ref{eq:tensor_decomposition}) of the amplitude is completely general and valid to all orders in perturbation theory, as it is based only on Lorentz symmetry and gauge invariance. 
  The dimensionality of the space-time is taken as $d = 4 + \epsilon$, where $\epsilon$ is the dimensional regularization parameter.
  
  The form factors $A_{1}$ and $A_{2}$ can be extracted from the amplitudes using appropriate projectors. 
  The projectors can be written as
  \begin{align}
   & P_{1}  = C_{1,1} T^{\dagger}_{1}  + C_{1,2} T^{\dagger}_{2}, 
   & P_{2}  = C_{2,1} T^{\dagger}_{1}  + C_{2,2} T^{\dagger}_{2}.  
  \end{align}
  The coefficients $C_{i,j}$ are determined by imposing the conditions :
  \begin{align}
    & \sum_{spins} P_{1} \mathcal{A} = A_{1}, 
    & \sum_{spins} P_{2} \mathcal{A} = A_{2}.
  \end{align}
  The Projectors are found to be,
  \begin{align}\label{eq:projectors}
    & P_{1}  = \frac{1}{stu(d-3)} \left[ \frac{2}{s}(d-2) T^{\dagger}_{1}  + T^{\dagger}_{2} \right],  
    & P_{2}  = \frac{1}{stu(d-3)} \left[ T^{\dagger}_{1} + \frac{s}{2} T^{\dagger}_{2} \right].
  \end{align} 
  The form-factors ($A_{i}$) can be perturbatively expanded in terms strong coupling constant \\($a_{s} = g_{s}^2/16 \pi^2 $) as,
    \begin{align}
    A_{i} = B_{f}\left[ A_{i}^{(0)} + a_{s} A_{i}^{(1)} + a_{s}^{2} A_{i}^{(2)} + \mathcal{O}(a_{s}^{3}) \right], \quad i=1,2.
  \end{align}
  Here $B_{f}$ is the overall Born factor defined as,
  \begin{align}
    B_{f} =  4 \pi \sqrt{a_{s}} ~ \lambda ~ (T^{a})_{j i},
  \end{align}
  where $\lambda = - i m_{b}/v$ is the Yukawa coupling of the Higgs boson to bottom quarks. 
  Here, $v$ is the vacuum expectation value, $m_{b}$ is the bottom-quark mass, and $(T^{a})_{j i}$ is the Gell-Mann matrix in the fundamental representation of the SU(3) colour group. 
  The form factors can be obtained by applying the projectors (\ref{eq:projectors}) to the amplitude, order by order in perturbation theory, as we discuss in the next section.


\section{Calculation}
We generate the Feynman diagrams for the process $H \rightarrow b + \bar{b} + g$ using \texttt{QGRAF}~\cite{Nogueira:1991ex}. 
At the tree level, two diagrams contribute to the amplitude. At one loop level, there are 13 diagrams, while at the two-loop level, there are 251 diagrams.
The form-factors $A_{1}$ and $A_{2}$ are obtained by applying the projectors (\ref{eq:projectors}) on the full amplitude and summing over the gluon polarization defined in eq. (\ref{eq:gluon_polarization_sum}).
We use our in-house-developed \texttt{FORM}~\cite{Ruijl:2017dtg} routines for Lorentz contractions and colour algebra, after which we obtain a large number of scalar integrals.
First, we use \texttt{Reduze}~\cite{Studerus:2009ye,vonManteuffel:2012np} to find relations for shifting the momenta to map the large number of integrals to two families 
of integrals, one planar ($F_{1}$) and one non-planar ($F_{2}$), defined in ~\cite{Gehrmann:2023etk}. 
We then use \texttt{Kira}~\cite{Lange:2025fba} to reduce the integrals to master integrals (MI) using integration by parts (IBP) and Lorentz invariance identities~\cite{Gehrmann:1999as} through the implementation of the Laporta algorithm ~\cite{Laporta:2000dsw}. 
Once the expressions for the form factors are obtained in terms of master integrals, we substitute the integrals that are known in the literature~\cite{Gehrmann:2023etk,Gehrmann:2001ck,Gehrmann:2000zt}. 
We also note that these master integrals are the same as the ones used in  ~\cite{Banerjee:2024zho} for the pseudoscalar decay processes $A \rightarrow g + g + g$ and $A \rightarrow q + \bar{q} + g$.


The master integrals are known in terms of the dimensionless variables $x$, $y$ and $z$ defined as,
\begin{align}\label{eq:xyz_definition}
x = \frac{s}{M_H^2}, \quad y = \frac{u}{M_H^2}, \quad z = \frac{t}{M_H^2},
\end{align}
with the constraint $x + y + z = 1$. 
The master integrals are polynomials in powers of $\epsilon$ with coefficients containing Goncharov polylogarithms (GPL).
The definition of these GPLs is given in appendix~\ref{appA}.
The results for the form factors are obtained up to $\mathcal{O}(\epsilon^{4})$ at one loop and up to $\mathcal{O}(\epsilon^{2})$ at the two-loop level. 
Once the unrenormalized form factors are obtained, we further simplify them using our in-house \texttt{Mathematica} simplification routines, alongside packages such as MultivariateApart~\cite{Heller:2021qkz}.

\subsection{UV Renormalization and IR subtraction}
The unrenormalized form factors contain ultraviolet (UV) and infrared (IR) divergences which appear as poles in $\epsilon(=d-4)$ in dimensional regularization with regularization scale $\mu_{0}$.
We have performed UV renormalization of the form factors in the modified minimal-subtraction ($\overline{MS}$) scheme. For the process under consideration, both the strong coupling constant and the Yukawa coupling of the Higgs boson to bottom quarks require renormalization. We follow the renormalization procedure as described in~\cite{Ahmed:2014pka}. For completeness, we also describe the renormalization procedure here.

The bare strong coupling constant $a^{b}_{s}$ is related to the renormalized strong coupling constant $a_{s} \equiv a_{s}(\mu^{2})$ at the renormalization scale $\mu$ as, 
\begin{align}\label{eq:renormalization_as}
  \frac{a_{s}^{b}}{\mu_{0}^{\epsilon}} S_{\epsilon} = \frac{a_{s}}{\mu^{\epsilon}} \left[ 1 
  + a_{s} Z_{1}
  + a_{s}^{2} Z_{2}
  + \mathcal{O}\left(a_{s}^{3}\right) \right]
\end{align}
where,
\begin{align}
  & Z_{1} =  \frac{2 \beta_{0}}{\epsilon}, 
  & Z_{2} =  \left( \frac{4 \beta_{0}^{2}}{\epsilon^{2}} + \frac{\beta_{1}}{\epsilon}\right). 
\end{align} 
The Yukawa coupling renormalization is given as,
\begin{align}\label{eq:renormalization_Yukawa}
  \frac{\lambda^{b}}{\mu_{0}^{\epsilon}} S_{\epsilon} = \frac{\lambda}{\mu^{\epsilon}} \left[ 1 
  + a_{s} Z_{\lambda,1}
  + a_{s}^{2} Z_{\lambda,2}
  + \mathcal{O}\left(a_{s}^{3}\right) \right]
\end{align}
where,
\begin{align}
  & Z_{\lambda,1} = \frac{1}{\epsilon} \left( 6 C_{F} \right),
  & Z_{\lambda,2} = \frac{1}{\epsilon^{2}} \left( 18 C_{F}^{2} - 6 \beta_{0} C_{F} \right) + \frac{1}{\epsilon} \left( \frac{3}{2} C_{F}^{2} + \frac{97}{6} C_{F} C_{A} - \frac{5}{3} n_{f} C_{F} \right) 
\end{align}

with $S_{\epsilon} = \exp\left[ \frac{\epsilon}{2} \left( \gamma_{E} - \ln 4 \pi \right) \right]$, where $\gamma_{E}$ is the Euler-Mascheroni constant. 
The $C_{F}$ and $C_{A}$ are the Casimir invariants of the SU(3) colour group in the fundamental and adjoint representation, $n_{f}$ is the number of active quark flavours. The coefficients of QCD beta function $\beta_{0}$ and $\beta_{1}$ are given as,
\begin{align}
  & \beta_{0} = \frac{11}{3} C_{A} - \frac{2}{3} n_{f},
  & \beta_{1} = \frac{34}{3} C_{A}^{2} - \frac{10}{3} C_{A} n_{f} - 2 C_{F} n_{f}. 
\end{align}
The renormalized form factors are obtained by substituting eqs. (\ref{eq:renormalization_as}) and (\ref{eq:renormalization_Yukawa}) in the unrenormalized form factors and re-expanding in powers of renormalized strong coupling constant $a_{s}$. The renormalized form factors at different orders in $a_{s}$ can be written as, 
\begin{align}
  A_{i}^{(0)} &= \left( \frac{1}{\mu^{\epsilon}} \right)^{\frac{1}{2}} A_{i}^{b,(0)}, \label{Eq:renormalized_formfactor_0}\\ 
 A_{i}^{(1)} &= \left( \frac{1}{\mu^{\epsilon}} \right)^{\frac{3}{2}}\left[ A_{i}^{b,(1)} + \frac{ \mu^{\epsilon} }{\epsilon}\left( \beta_{0} + 6 C_{F} \right)A_{i}^{b,(0)}\right], \label{Eq:renormalized_formfactor_1}\\  
  A_{i}^{(2)} &= \left( \frac{1}{\mu^{\epsilon}} \right)^{\frac{5}{2}} \bigg[
    A_{i}^{b,(2)}   
  + \frac{ \mu^{\epsilon } }{\epsilon} \left(3 \beta_{0} + 6 C_{F} \right)A_{b}^{b,(1)}  \nonumber \\
  &~~~+ \frac{\mu^{2 \epsilon }}{\epsilon^2}\left\{\frac{3}{2}\beta_{0} + 12 \beta_{0}C_{F} + 18 C_{F}^2 + \epsilon \left( \frac{\beta_{1}}{2} + \frac{97}{6}C_{A} C_{F} + \frac{3}{2}C_{F}^2 - \frac{5}{3}C_{F}n_{f}\right)   \right\}A_{i}^{b,(0)} \bigg]. \label{Eq:renormalized_formfactor_2}
\end{align}

After UV renormalization, the form factors still contain IR divergences as poles in $\epsilon$. In any IR-safe observable, these IR divergences will cancel against those coming from the real emission contributions as per the KLN theorem~\cite{Kinoshita:1962ur,Lee:1964is}.
The structure of these IR divergences is universal and is well understood~\cite{Catani_1998}. We follow the subtraction procedure as described in~\cite{Gehrmann:2023etk, Ahmed:2014pka} to obtain the finite form factors. The renormalized form factors can be written in terms of finite form factors as,
\begin{align}
  A_{i}^{(1)} &= 2\mathbf{I}_{b}^{(1)}(\epsilon) A_{i}^{(0)} + A_{i}^{(1),fin}, \\
  A_{i}^{(2)} &= 4 \mathbf{I}_{b}^{(2)}(\epsilon) A_{i}^{(0)} + 2 \mathbf{I}_{b}^{(1)}(\epsilon) A_{i}^{(1)} + A_{i}^{(2),fin},
\end{align}
where the operators $\mathbf{I}_{b}^{(1)}(\epsilon)$ and $\mathbf{I}_{b}^{(2)}(\epsilon)$ are given as,
\begin{align}
  \mathbf{I}_{b}^{(1)}(\epsilon) = &\frac{1}{2}\frac{\exp[-\frac{\epsilon}{2} \gamma_{E}]}{\Gamma(1 + \frac{\epsilon}{2})} 
  \Bigg[
     \left( \frac{4}{\epsilon^2} - \frac{3}{\epsilon} \right) \left( C_{A} - 2 C_{F} \right) \left( \frac{\mu^{2}}{s} \right)^{-\frac{\epsilon}{2}} 
     \nonumber \\
     &
     + \left( -\frac{4}{\epsilon^2}C_{A} + \frac{3}{2 \epsilon}C_{A} + \frac{\beta_{0}}{2 \epsilon} \right) \left\{  \left(\frac{\mu^{2}}{t}\right)^{-\frac{\epsilon}{2}  } + \left(\frac{\mu^{2}}{u}\right)^{-\frac{\epsilon}{2}}
     \right\} 
  \Bigg],  \\
  \mathbf{I}_{b}^{(2)}(\epsilon) = &-\frac{1}{2} \mathbf{I}_{b}^{(1)}(\epsilon) \left( \mathbf{I}_{b}^{(1)}(\epsilon) - 2 \frac{\beta_{0}}{\epsilon} \right)
  + \frac{\exp[{\frac{\epsilon}{2} \gamma_{E} }]}{\Gamma\left(1 + \frac{\epsilon}{2} \right)  } \left(-\frac{\beta_{0}}{\epsilon} + \mathbf{K} \right) \mathbf{I}_{b}^{(1)}(2 \epsilon) 
  \nonumber \\ 
  &
  + \left( 2~\mathbf{H}_{q}^{(2)}(\epsilon) + \mathbf{H}_{g}^{(2)}(\epsilon) \right), 
  \end{align}
  where,
  \begin{align}
    &\mathbf{K} = \left( \frac{67}{18} - \frac{\pi^{2}}{6} \right) C_{A} - \frac{5}{9} n_{f},
    \end{align}
  \begin{align}
    &\mathbf{H}_{q}^{(2)} = \frac{1}{\epsilon} \Bigg[ C_{A}C_{F} \left( \frac{245}{432} + \frac{23}{16} \zeta_{2} - \frac{13}{4}\zeta_{3} \right)
    + C_{F}^{2} \left( \frac{3}{16} - \frac{3}{2} \zeta_{3} + 3 \zeta_{3} \right) + C_{F} n_{f} \left( \frac{25}{216} - \frac{1}{8}\zeta_{2} \right) 
      \Bigg], 
    \end{align}
  \begin{align}
    &\mathbf{H}_{g}^{(2)} = \frac{1}{\epsilon} \Bigg[ 
      C_{A}^{2} \left( -\frac{5}{24} - \frac{11}{48} \zeta_{2} - \frac{1}{4}\zeta_{3}  \right)
      + C_{A} n_{f} \left( \frac{29}{54} + \frac{1}{24}\zeta_{2} \right)
      - \frac{1}{4}C_{F} n_{f} 
      - \frac{5}{54} n_{f}^{2}
      \Bigg].
    \end{align}

  In order to check the correctness of our results. 
  We verify that the renormalized form factors reproduce the expected IR structure as predicted by the operators $\mathbf{I}_{b}^{(1)}(\epsilon)$ and $\mathbf{I}_{b}^{(2)}(\epsilon)$. 
  We have performed checks at both the one-loop and two-loop levels, with results from independent calculations available in the literature~\cite{Ahmed:2014pka,Mondini:2019vub}. 
  We found them completely agreeing with he results up to $\mathcal{O}(\epsilon^{0})$. 
  The squared amplitude (see appendix~\ref{appB} for details) can be easily evaluated using these form-factors.



\begin{figure}[ht!]
\centering
\includegraphics[width=\textwidth]{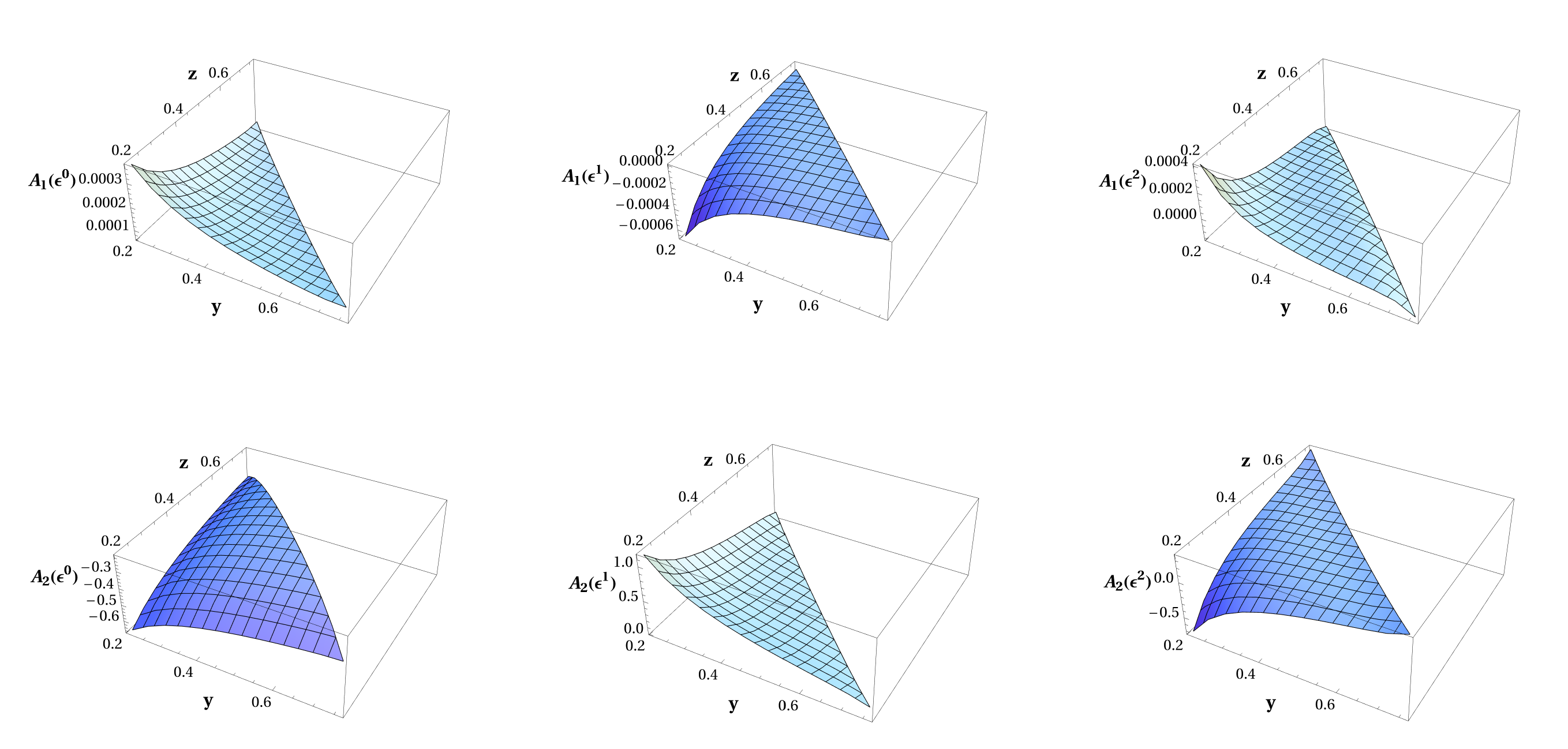}
\caption{Two-loop form factors ($A_{1}$ and $A_{2}$) in different powers of dimensional regulator $\epsilon$ in the region $0.2 < y < 0.8$ and $0.2 < z < 0.8$.}
\label{fig:form_factor}
\end{figure}
\begin{figure}[ht!]
\centering
\includegraphics[width=\textwidth]{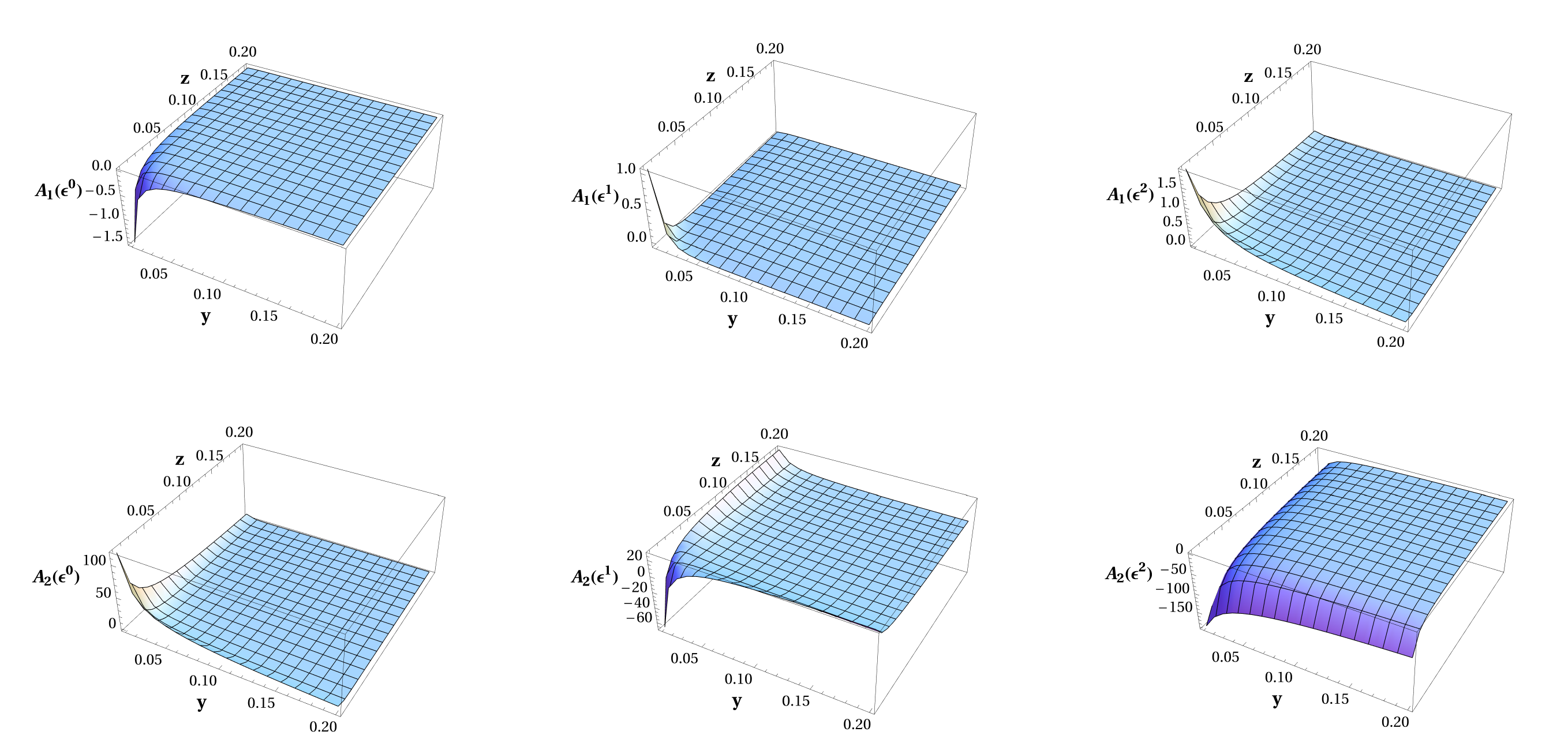}
\caption{Two-loop form factors ($A_{1}$ and $A_{2}$) in different powers of dimensional regulator $\epsilon$ in the region $0.01 < y  < 0.2 $ and $ 0.01 < z < 0.2$.}
\label{fig:form_factor2}
\end{figure}

In what follows, we show the numerical behaviour of the two-loop form-factors $A_{1}$ and $A_{2}$ up to $\epsilon^2$. 
The numerical results are evaluated with
\begin{align}
   & N=3, 
   \quad \quad
   \mu = Q = 125 ~\mathrm{GeV} 
   \quad
   \mathrm{and}
   \quad \quad 
    n_f = 5.  \nonumber 
\end{align}
These form-factors are finite in the region $0<y<1$, $0<z<1-y$. However, they will give large contributions in the limit $y \to 0$ or $z \to 0$. 
In figure~\ref{fig:form_factor} we show the behaviour of the two-loop form-factor $A_{1}$ ($A_{2}$) in the upper (lower) panel in the region $0.2 < y < 1$ and $0.2 < z < 1$.
The time taken for evaluations of the form-factors per phase-space point is given in table~\ref{tab:timings}.
\input{timing.tex}
The computation has been done on a 3.2 GHz computer with 64Gb RAM by implementing the form factors in a FORTRAN-95 code and using \texttt{handyg}~\cite{Naterop:2019xaf} for evaluating the GPLs.
As can be seen, the computational power required for higher powers of $\epsilon$ terms increases by an order of magnitude compared to lower powers.
This is due to the presence of higher-weight GPLs in the coefficients of higher powers of $\epsilon$, as also observed in \cite{Banerjee:2024zho}. 
In figure~\ref{fig:form_factor2}, we show the same form-factors as in figure~\ref{fig:form_factor}, but in the region $0.01 < y < 0.2$ and $0.01 < z < 0.2$.

\section{Summary}
In this work, we computed the two-loop massless QCD correction to the Higgs boson decay to a bottom quark pair in association with a gluon ($H \rightarrow b + \bar{b} + g$) at higher powers of the dimensional regularization parameter $\epsilon$.
We have used the projection method to extract the form factors from the amplitude. Our results of these amplitudes are crucial to construct N$^{3}$LO QCD correction to the Higgs boson production in the bottom anti-bottom annihilation process in association with a jet. 
We have confirmed that our results reproduce the expected infrared structure as predicted by Catani's IR subtraction operators. 
We observe that due to the presence of the higher weight GPLs in the coefficients of $\epsilon^2$, the computational time for the form-factors rapidly increases compared to that of $\epsilon^0$.
The two-loop amplitudes computed in this work provide essential ingredients for future high-precision (three-loop) calculations of Higgs boson production in bottom quark-initiated channels at hadron colliders.

\section*{Acknowledgements}
The work of C.D. at Physical Research Laboratory was supported by Department of Space, Govt. of India.

\begin{appendix}
\numberwithin{equation}{section}

\section{GPLs}\label{appA}

We recall here that the GPLs are defined as iterated integrals over the rational functions as,
\begin{align}
    G(l_1,l_2,...,l_n;x) = &\int_{0}^{x}\frac{dt}{t-l_1} ~G(l_2,...,l_n;t),\\
     G(\underbrace{0,0,..,0}_n;x) = & \frac{1}{n!}\text{ln}^n(x),~ G(x) = 1;
\end{align}

We write few weight-1 GPLs in terms of logarithms~\cite{Duhr:2019tlz},
\begin{eqnarray}
G(1;y) & = & \ln (1-y)\; ,\nonumber \\
G(0;y) & = & \ln y \; ,\nonumber \\
G(-1;y) & = & \ln (1+y) \nonumber \\
G(1-z;y) &=&  \ln\left(1-\frac{y}{1-z}\right) \ , \nonumber \\
G(z;y) &=&  \ln\left(\frac{z-y}{z}\right) .
\end{eqnarray}
The derivative of the GPLs can be written as,
\begin{align}
  \frac{ \partial }{ \partial y} G(a,\vec{b};y) = f(a,y) G(\vec{b};y)
\end{align}
Where, $f(a,y) = \frac{1}{a-y}$.

\section{Squared amplitude}\label{appB}

In order to calculate any physical observable, we need the squared amplitude $|\mathcal{A}|^2$. This can be easily calculated following tensor decomposition of the amplitude \ref{eq:tensor_decomposition}. Using relations ~\ref{eq:stu} we get,

\begin{align}
  \left|\mathcal{A}\right|^{2} =\frac{1}{2} A_{1}^{\dag} A_1 ~ s^2tu - \left(A_1^{\dag}A_2 + A_2^{\dag}A_1\right) ~ stu + A_2^{\dag}A_2(2 d - 4) ~ tu ,
\end{align}
where $d$ is the number of dimensions.

\end{appendix}

\pagebreak

\bibliographystyle{JHEP}
\bibliography{references.bib}


\end{document}

%% file: timing.tex
\begin{table}[ht!]
\centering
\renewcommand{\arraystretch}{1.3}
\begin{tabular}{|c|c|}
\hline
\textbf{$\mathbf{\epsilon}$ order} & \textbf{Runtime (s)} \\
\hline
$\epsilon^0$ & 2.0 \\
\hline
$\epsilon^1$ & 16.0 \\
\hline
$\epsilon^2$ & 266.0 \\
\hline
\end{tabular}
\caption{Numerical evaluation time for the form-factors ($A_{1},A_{2}$) at different powers of dimensional regulator $\epsilon$ for a choice of parameters $y = 0.34$ and $z = 0.56$.}
\label{tab:timings}
\end{table}

%% file: references.bib
@article{ATLAS:2012yve,
    author = "Aad, Georges and others",
    collaboration = "ATLAS",
    title = "{Observation of a new particle in the search for the Standard Model Higgs boson with the ATLAS detector at the LHC}",
    eprint = "1207.7214",
    archivePrefix = "arXiv",
    primaryClass = "hep-ex",
    reportNumber = "CERN-PH-EP-2012-218",
    doi = "10.1016/j.physletb.2012.08.020",
    journal = "Phys. Lett. B",
    volume = "716",
    pages = "1--29",
    year = "2012"
}

@article{Campbell:2019gmd,
    author = "Campbell, John M. and Ellis, R. Keith and Seth, Satyajit",
    title = "{H + 1 jet production revisited}",
    eprint = "1906.01020",
    archivePrefix = "arXiv",
    primaryClass = "hep-ph",
    reportNumber = "FERMILAB-PUB-19-189-T, IPPP/19/33",
    doi = "10.1007/JHEP10(2019)136",
    journal = "JHEP",
    volume = "10",
    pages = "136",
    year = "2019"
}

@article{Chen:2019wxf,
    author = "Chen, X. and Gehrmann, T. and Glover, E. W. N. and Huss, A.",
    title = "{Fiducial cross sections for the four-lepton decay mode in Higgs-plus-jet production up to NNLO QCD}",
    eprint = "1905.13738",
    archivePrefix = "arXiv",
    primaryClass = "hep-ph",
    reportNumber = "ZU-TH 27/19, IPPP/19/45, CERN-TH-2019-083",
    doi = "10.1007/JHEP07(2019)052",
    journal = "JHEP",
    volume = "07",
    pages = "052",
    year = "2019"
}

@article{Chen:2016zka,
    author = "Chen, X. and Cruz-Martinez, J. and Gehrmann, T. and Glover, E. W. N. and Jaquier, M.",
    title = "{NNLO QCD corrections to Higgs boson production at large transverse momentum}",
    eprint = "1607.08817",
    archivePrefix = "arXiv",
    primaryClass = "hep-ph",
    reportNumber = "IPPP-16-73, ZU-TH-29-16, FR-PHENO-2016-014",
    doi = "10.1007/JHEP10(2016)066",
    journal = "JHEP",
    volume = "10",
    pages = "066",
    year = "2016"
}

@article{Chen:2018pzu,
    author = "Chen, Xuan and Gehrmann, Thomas and Glover, E. W. Nigel and Huss, Alexander and Li, Ye and Neill, Duff and Schulze, Markus and Stewart, Iain W. and Zhu, Hua Xing",
    title = "{Precise QCD Description of the Higgs Boson Transverse Momentum Spectrum}",
    eprint = "1805.00736",
    archivePrefix = "arXiv",
    primaryClass = "hep-ph",
    reportNumber = "IPPP/18/28, ZU-TH 16/18, CERN-TH-2018-108, MIT-CTP 5013, HU-EP-18/14, FERMILAB-PUB-18-212-T",
    doi = "10.1016/j.physletb.2018.11.037",
    journal = "Phys. Lett. B",
    volume = "788",
    pages = "425--430",
    year = "2019"
}

@article{Chen:2014gva,
    author = "Chen, X. and Gehrmann, T. and Glover, E. W. N. and Jaquier, M.",
    title = "{Precise QCD predictions for the production of Higgs + jet final states}",
    eprint = "1408.5325",
    archivePrefix = "arXiv",
    primaryClass = "hep-ph",
    reportNumber = "IPPP-14-64, ZU-TH-27-14",
    doi = "10.1016/j.physletb.2014.11.021",
    journal = "Phys. Lett. B",
    volume = "740",
    pages = "147--150",
    year = "2015"
}

@article{Boughezal:2015dra,
    author = "Boughezal, Radja and Caola, Fabrizio and Melnikov, Kirill and Petriello, Frank and Schulze, Markus",
    title = "{Higgs boson production in association with a jet at next-to-next-to-leading order}",
    eprint = "1504.07922",
    archivePrefix = "arXiv",
    primaryClass = "hep-ph",
    reportNumber = "CERN-PH-TH-2015-056, TTP15-017",
    doi = "10.1103/PhysRevLett.115.082003",
    journal = "Phys. Rev. Lett.",
    volume = "115",
    number = "8",
    pages = "082003",
    year = "2015"
}

@article{Buonocore:2025fqs,
    author = "Buonocore, Luca and Grazzini, Massimiliano and Kallweit, Stefan and Lindert, Jonas M. and Savoini, Chiara",
    title = "{Towards NNLO QCD predictions for off-shell top-quark pair production and decays}",
    eprint = "2507.11410",
    archivePrefix = "arXiv",
    primaryClass = "hep-ph",
    reportNumber = "ZU-TH-47/25, TUM-HEP-1565/25, CERN-TH-2025-130",
    doi = "10.1007/JHEP10(2025)195",
    journal = "JHEP",
    volume = "10",
    pages = "195",
    year = "2025"
}

@article{Agarwal:2024jyq,
    author = "Agarwal, Bakul and Heinrich, Gudrun and Jones, Stephen P. and Kerner, Matthias and Klein, Sven Yannick and Lang, Jannis and Magerya, Vitaly and Olsson, Anton",
    title = "{Two-loop amplitudes for $ t\overline{t}H $ production: the quark-initiated N$_{f}$-part}",
    eprint = "2402.03301",
    archivePrefix = "arXiv",
    primaryClass = "hep-ph",
    reportNumber = "IPPP/24/03, KA-TP-02-2024, P3H-24-007, TTK-24-03",
    doi = "10.1007/JHEP05(2024)013",
    journal = "JHEP",
    volume = "05",
    pages = "013",
    year = "2024",
    note = "[Erratum: JHEP 06, 142 (2024)]"
}

@article{Devoto:2024nhl,
    author = "Devoto, Simone and Grazzini, Massimiliano and Kallweit, Stefan and Mazzitelli, Javier and Savoini, Chiara",
    title = "{Precise predictions for $ t\overline{t}H $ production at the LHC: inclusive cross section and differential distributions}",
    eprint = "2411.15340",
    archivePrefix = "arXiv",
    primaryClass = "hep-ph",
    reportNumber = "ZU-TH-57/24, PSI-PR-24-22, TUM-HEP-1538/24",
    doi = "10.1007/JHEP03(2025)189",
    journal = "JHEP",
    volume = "03",
    pages = "189",
    year = "2025"
}

@article{CMS:2021sdq,
    author = "Tumasyan, Armen and others",
    collaboration = "CMS",
    title = "{Analysis of the $CP$ structure of the Yukawa coupling between the Higgs boson and $\tau$ leptons in proton-proton collisions at $ \sqrt{s} $ = 13 TeV}",
    eprint = "2110.04836",
    archivePrefix = "arXiv",
    primaryClass = "hep-ex",
    reportNumber = "CMS-HIG-20-006, CERN-EP-2021-189",
    doi = "10.1007/JHEP06(2022)012",
    journal = "JHEP",
    volume = "06",
    pages = "012",
    year = "2022"
}

@article{CMS:2020cga,
    author = "Sirunyan, Albert M and others",
    collaboration = "CMS",
    title = "{Measurements of $\mathrm{t\bar{t}}H$ Production and the CP Structure of the Yukawa Interaction between the Higgs Boson and Top Quark in the Diphoton Decay Channel}",
    eprint = "2003.10866",
    archivePrefix = "arXiv",
    primaryClass = "hep-ex",
    reportNumber = "CMS-HIG-19-013, CERN-EP-2020-028",
    doi = "10.1103/PhysRevLett.125.061801",
    journal = "Phys. Rev. Lett.",
    volume = "125",
    number = "6",
    pages = "061801",
    year = "2020"
}

@article{CMS:2014nkk,
    author = "Khachatryan, Vardan and others",
    collaboration = "CMS",
    title = "{Constraints on the spin-parity and anomalous HVV couplings of the Higgs boson in proton collisions at 7 and 8 TeV}",
    eprint = "1411.3441",
    archivePrefix = "arXiv",
    primaryClass = "hep-ex",
    reportNumber = "CMS-HIG-14-018, CERN-PH-EP-2014-265",
    doi = "10.1103/PhysRevD.92.012004",
    journal = "Phys. Rev. D",
    volume = "92",
    number = "1",
    pages = "012004",
    year = "2015"
}

@article{ATLAS:2013xga,
    author = "Aad, Georges and others",
    collaboration = "ATLAS",
    title = "{Evidence for the spin-0 nature of the Higgs boson using ATLAS data}",
    eprint = "1307.1432",
    archivePrefix = "arXiv",
    primaryClass = "hep-ex",
    reportNumber = "CERN-PH-EP-2013-102",
    doi = "10.1016/j.physletb.2013.08.026",
    journal = "Phys. Lett. B",
    volume = "726",
    pages = "120--144",
    year = "2013"
}

@article{ATLAS:2020rej,
    author = "Aad, Georges and others",
    collaboration = "ATLAS",
    title = "{Higgs boson production cross-section measurements and their EFT interpretation in the $4\ell $ decay channel at $\sqrt{s}=$13 TeV with the ATLAS detector}",
    eprint = "2004.03447",
    archivePrefix = "arXiv",
    primaryClass = "hep-ex",
    reportNumber = "CERN-EP-2020-034",
    doi = "10.1140/epjc/s10052-020-8227-9",
    journal = "Eur. Phys. J. C",
    volume = "80",
    number = "10",
    pages = "957",
    year = "2020",
    note = "[Erratum: Eur.Phys.J.C 81, 29 (2021), Erratum: Eur.Phys.J.C 81, 398 (2021)]"
}

@article{Catani:2021cbl,
    author = "Catani, Stefano and Fabre, Ignacio and Grazzini, Massimiliano and Kallweit, Stefan",
    title = "{${t {{\bar{t}}}H}$ production at NNLO: the flavour off-diagonal channels}",
    eprint = "2102.03256",
    archivePrefix = "arXiv",
    primaryClass = "hep-ph",
    reportNumber = "ZU-TH 3/21, ICAS 061/21",
    doi = "10.1140/epjc/s10052-021-09247-w",
    journal = "Eur. Phys. J. C",
    volume = "81",
    number = "6",
    pages = "491",
    year = "2021"
}

@article{Duhr:2019kwi,
    author = "Duhr, Claude and Dulat, Falko and Mistlberger, Bernhard",
    title = "{Higgs Boson Production in Bottom-Quark Fusion to Third Order in the Strong Coupling}",
    eprint = "1904.09990",
    archivePrefix = "arXiv",
    primaryClass = "hep-ph",
    reportNumber = "CERN-TH-2019-052,CP3-19-19,MIT-CTP/5115,SLAC-PUB-17425, CERN-TH-2019-052, CP3-19-19, MIT-CTP/5115, SLAC-PUB-17425",
    doi = "10.1103/PhysRevLett.125.051804",
    journal = "Phys. Rev. Lett.",
    volume = "125",
    number = "5",
    pages = "051804",
    year = "2020"
}

@article{CMS:2012qbp,
    author = "Chatrchyan, Serguei and others",
    collaboration = "CMS",
    title = "{Observation of a New Boson at a Mass of 125 GeV with the CMS Experiment at the LHC}",
    eprint = "1207.7235",
    archivePrefix = "arXiv",
    primaryClass = "hep-ex",
    reportNumber = "CMS-HIG-12-028, CERN-PH-EP-2012-220",
    doi = "10.1016/j.physletb.2012.08.021",
    journal = "Phys. Lett. B",
    volume = "716",
    pages = "30--61",
    year = "2012"
}

@article{Boughezal:2015aha,
    author = "Boughezal, Radja and Focke, Christfried and Giele, Walter and Liu, Xiaohui and Petriello, Frank",
    title = "{Higgs boson production in association with a jet at NNLO using jettiness subtraction}",
    eprint = "1505.03893",
    archivePrefix = "arXiv",
    primaryClass = "hep-ph",
    reportNumber = "FERMILAB-PUB-15-210-T",
    doi = "10.1016/j.physletb.2015.06.055",
    journal = "Phys. Lett. B",
    volume = "748",
    pages = "5--8",
    year = "2015"
}

@article{Higgs:1964ia,
    author = "Higgs, Peter W.",
    title = "{Broken symmetries, massless particles and gauge fields}",
    doi = "10.1016/0031-9163(64)91136-9",
    journal = "Phys. Lett.",
    volume = "12",
    pages = "132--133",
    year = "1964"
}

@article{PhysRevLett.13.508,
  title = {Broken Symmetries and the Masses of Gauge Bosons},
  author = {Higgs, Peter W.},
  journal = {Phys. Rev. Lett.},
  volume = {13},
  issue = {16},
  pages = {508--509},
  numpages = {0},
  year = {1964},
  month = {Oct},
  publisher = {American Physical Society},
  doi = {10.1103/PhysRevLett.13.508},
  url = {https://link.aps.org/doi/10.1103/PhysRevLett.13.508}
}

@article{Higgs:1966ev,
    author = "Higgs, Peter W.",
    title = "{Spontaneous Symmetry Breakdown without Massless Bosons}",
    doi = "10.1103/PhysRev.145.1156",
    journal = "Phys. Rev.",
    volume = "145",
    pages = "1156--1163",
    year = "1966"
}

@article{Englert:1964et,
    author = "Englert, F. and Brout, R.",
    editor = "Taylor, J. C.",
    title = "{Broken Symmetry and the Mass of Gauge Vector Mesons}",
    doi = "10.1103/PhysRevLett.13.321",
    journal = "Phys. Rev. Lett.",
    volume = "13",
    pages = "321--323",
    year = "1964"
}

@article{Guralnik:1964eu,
    author = "Guralnik, G. S. and Hagen, C. R. and Kibble, T. W. B.",
    editor = "Taylor, J. C.",
    title = "{Global Conservation Laws and Massless Particles}",
    doi = "10.1103/PhysRevLett.13.585",
    journal = "Phys. Rev. Lett.",
    volume = "13",
    pages = "585--587",
    year = "1964"
}

@article{Ahmed:2014pka,
    author = "Ahmed, Taushif and Mahakhud, Maguni and Mathews, Prakash and Rana, Narayan and Ravindran, V.",
    title = "{Two-loop QCD corrections to Higgs $\to b+\overline{b}+g$ amplitude}",
    eprint = "1405.2324",
    archivePrefix = "arXiv",
    primaryClass = "hep-ph",
    reportNumber = "HRI-RECAPP-2014-010",
    doi = "10.1007/JHEP08(2014)075",
    journal = "JHEP",
    volume = "08",
    pages = "075",
    year = "2014"
}

@article{Mondini:2021nck,
    author = "Mondini, Roberto and Williams, Ciaran",
    title = "{Bottom-induced contributions to Higgs plus jet at next-to-next-to-leading order}",
    eprint = "2102.05487",
    archivePrefix = "arXiv",
    primaryClass = "hep-ph",
    doi = "10.1007/JHEP05(2021)045",
    journal = "JHEP",
    volume = "05",
    pages = "045",
    year = "2021"
}

@article{Mondini:2019vub,
    author = "Mondini, Roberto and Williams, Ciaran",
    title = "{$ H\to b\overline{b}j $ at next-to-next-to-leading order accuracy}",
    eprint = "1904.08961",
    archivePrefix = "arXiv",
    primaryClass = "hep-ph",
    doi = "10.1007/JHEP06(2019)120",
    journal = "JHEP",
    volume = "06",
    pages = "120",
    year = "2019"
}

@article{Banerjee:2017faz,
    author = "Banerjee, Pulak and Dhani, Prasanna K. and Ravindran, V.",
    title = "{Two loop QCD corrections for the process Pseudo-scalar Higgs $\rightarrow 3$ partons}",
    eprint = "1708.02387",
    archivePrefix = "arXiv",
    primaryClass = "hep-ph",
    reportNumber = "IMSC-2017-08-06, IMSc/2017/08/06",
    doi = "10.1007/JHEP10(2017)067",
    journal = "JHEP",
    volume = "10",
    pages = "067",
    year = "2017"
}

@article{Cal:2023mib,
    author = "Cal, Pedro and von Kuk, Rebecca and Lim, Matthew A. and Tackmann, Frank J.",
    title = "{qT spectrum for Higgs boson production via heavy quark annihilation at N3LL'+aN3LO}",
    eprint = "2306.16458",
    archivePrefix = "arXiv",
    primaryClass = "hep-ph",
    doi = "10.1103/PhysRevD.110.076005",
    journal = "Phys. Rev. D",
    volume = "110",
    number = "7",
    pages = "076005",
    year = "2024"
}

@article{Dreyer:2016oyx,
    author = "Dreyer, Fr{\'e}d{\'e}ric A. and Karlberg, Alexander",
    title = "{Vector-Boson Fusion Higgs Production at Three Loops in QCD}",
    eprint = "1606.00840",
    archivePrefix = "arXiv",
    primaryClass = "hep-ph",
    reportNumber = "CERN-TH-2016-127, OUTP-16-12P",
    doi = "10.1103/PhysRevLett.117.072001",
    journal = "Phys. Rev. Lett.",
    volume = "117",
    number = "7",
    pages = "072001",
    year = "2016"
}

@article{Anastasiou:2015vya,
    author = "Anastasiou, Charalampos and Duhr, Claude and Dulat, Falko and Herzog, Franz and Mistlberger, Bernhard",
    title = "{Higgs Boson Gluon-Fusion Production in QCD at Three Loops}",
    eprint = "1503.06056",
    archivePrefix = "arXiv",
    primaryClass = "hep-ph",
    reportNumber = "CERN-PH-TH-2015-055, CP3-15-07",
    doi = "10.1103/PhysRevLett.114.212001",
    journal = "Phys. Rev. Lett.",
    volume = "114",
    pages = "212001",
    year = "2015"
}

@article{Baglio:2022wzu,
    author = "Baglio, Julien and Duhr, Claude and Mistlberger, Bernhard and Szafron, Robert",
    title = "{Inclusive production cross sections at N$^{3}$LO}",
    eprint = "2209.06138",
    archivePrefix = "arXiv",
    primaryClass = "hep-ph",
    reportNumber = "CERN-TH-2022-109, SLAC-PUB-17699, BONN-TH-2022-22",
    doi = "10.1007/JHEP12(2022)066",
    journal = "JHEP",
    volume = "12",
    pages = "066",
    year = "2022"
}

@article{Das:2022zie,
    author = "Das, Goutam and Dey, Chinmoy and Kumar, M. C. and Samanta, Kajal",
    title = "{Threshold enhanced cross sections for colorless productions}",
    eprint = "2210.17534",
    archivePrefix = "arXiv",
    primaryClass = "hep-ph",
    reportNumber = "TTK-22-34, P3H-22-106",
    doi = "10.1103/PhysRevD.107.034038",
    journal = "Phys. Rev. D",
    volume = "107",
    number = "3",
    pages = "034038",
    year = "2023"
}

@article{Gehrmann:2023etk,
    author = "Gehrmann, Thomas and Jakub{\v{c}}{\'\i}k, Petr and Mella, Cesare Carlo and Syrrakos, Nikolaos and Tancredi, Lorenzo",
    title = "{Two-loop helicity amplitudes for $H+$jet production to higher orders in the dimensional regulator}",
    eprint = "2301.10849",
    archivePrefix = "arXiv",
    primaryClass = "hep-ph",
    reportNumber = "ZU-TH 07/23, TUM-HEP 1451/23",
    doi = "10.1007/JHEP04(2023)016",
    journal = "JHEP",
    volume = "04",
    pages = "016",
    year = "2023"
}

@article{Gehrmann:2001ck,
    author = "Gehrmann, T. and Remiddi, E.",
    title = "{Two loop master integrals for gamma* --{\ensuremath{>}} 3 jets: The Nonplanar topologies}",
    eprint = "hep-ph/0101124",
    archivePrefix = "arXiv",
    reportNumber = "CERN-TH-2001-005",
    doi = "10.1016/S0550-3213(01)00074-8",
    journal = "Nucl. Phys. B",
    volume = "601",
    pages = "287--317",
    year = "2001"
}

@article{Gehrmann:2000zt,
    author = "Gehrmann, T. and Remiddi, E.",
    title = "{Two loop master integrals for gamma* ---{\ensuremath{>}} 3 jets: The Planar topologies}",
    eprint = "hep-ph/0008287",
    archivePrefix = "arXiv",
    reportNumber = "KA-TTP-00-20",
    doi = "10.1016/S0550-3213(01)00057-8",
    journal = "Nucl. Phys. B",
    volume = "601",
    pages = "248--286",
    year = "2001"
}

@article{Naterop:2019xaf,
    author = "Naterop, L. and Signer, A. and Ulrich, Y.",
    title = "{handyG {\textemdash}Rapid numerical evaluation of generalised polylogarithms in Fortran}",
    eprint = "1909.01656",
    archivePrefix = "arXiv",
    primaryClass = "hep-ph",
    reportNumber = "PSI-PR-19-17, ZU-TH 40/19",
    doi = "10.1016/j.cpc.2020.107165",
    journal = "Comput. Phys. Commun.",
    volume = "253",
    pages = "107165",
    year = "2020"
}

@article{Harlander:2003ai,
    author = "Harlander, Robert V. and Kilgore, William B.",
    title = "{Higgs boson production in bottom quark fusion at next-to-next-to leading order}",
    eprint = "hep-ph/0304035",
    archivePrefix = "arXiv",
    reportNumber = "BNL-HET-03-4, CERN-TH-2003-067",
    doi = "10.1103/PhysRevD.68.013001",
    journal = "Phys. Rev. D",
    volume = "68",
    pages = "013001",
    year = "2003"
}

@article{Banerjee:2024zho,
    author = "Banerjee, Pulak and Dey, Chinmoy and Kumar, M. C. and Ravindran, V.",
    title = "{Pseudoscalar Higgs boson decay to three parton amplitudes at NNLO to higher orders in the dimensional regulator}",
    eprint = "2411.17611",
    archivePrefix = "arXiv",
    primaryClass = "hep-ph",
    doi = "10.1103/PhysRevD.111.054037",
    journal = "Phys. Rev. D",
    volume = "111",
    number = "5",
    pages = "054037",
    year = "2025"
}

@inproceedings{Caola:2022ayt,
    author = "Caola, Fabrizio and Chen, Wen and Duhr, Claude and Liu, Xiaohui and Mistlberger, Bernhard and Petriello, Frank and Vita, Gherardo and Weinzierl, Stefan",
    title = "{The Path forward to N$^3$LO}",
    booktitle = "{Snowmass 2021}",
    eprint = "2203.06730",
    archivePrefix = "arXiv",
    primaryClass = "hep-ph",
    reportNumber = "SLAC-PUB-17658, BONN-TH-2022-06, MITP-22-021",
    month = "3",
    year = "2022"
}

@article{CMS:2024eka,
    author = "Hayrapetyan, Aram and others",
    collaboration = "CMS",
    title = "{Measurement of the Higgs boson mass and width using the four-lepton final state in proton-proton collisions at s=13{\,}{\,}TeV}",
    eprint = "2409.13663",
    archivePrefix = "arXiv",
    primaryClass = "hep-ex",
    reportNumber = "CMS-HIG-21-019, CERN-EP-2024-210",
    doi = "10.1103/PhysRevD.111.092014",
    journal = "Phys. Rev. D",
    volume = "111",
    number = "9",
    pages = "092014",
    year = "2025"
}

@article{Ahmed:2014cha,
    author = "Ahmed, Taushif and Rana, Narayan and Ravindran, V.",
    title = "{Higgs boson production through $b \bar b$ annihilation at threshold in N$^3$LO QCD}",
    eprint = "1408.0787",
    archivePrefix = "arXiv",
    primaryClass = "hep-ph",
    reportNumber = "HRI-RECAPP-2014-018",
    doi = "10.1007/JHEP10(2014)139",
    journal = "JHEP",
    volume = "10",
    pages = "139",
    year = "2014"
}

@article{ATLAS:2014yga,
    author = "Aad, Georges and others",
    collaboration = "ATLAS",
    title = "{Measurements of fiducial and differential cross sections for Higgs boson production in the diphoton decay channel at $\sqrt{s}=8$ TeV with ATLAS}",
    eprint = "1407.4222",
    archivePrefix = "arXiv",
    primaryClass = "hep-ex",
    reportNumber = "CERN-PH-EP-2014-148",
    doi = "10.1007/JHEP09(2014)112",
    journal = "JHEP",
    volume = "09",
    pages = "112",
    year = "2014"
}

@article{CMS:2015qgt,
    author = "Khachatryan, Vardan and others",
    collaboration = "CMS",
    title = "{Measurement of differential cross sections for Higgs boson production in the diphoton decay channel in pp collisions at $\sqrt{s}=8\,\text {TeV} $}",
    eprint = "1508.07819",
    archivePrefix = "arXiv",
    primaryClass = "hep-ex",
    reportNumber = "CMS-HIG-14-016, CERN-PH-EP-2015-195",
    doi = "10.1140/epjc/s10052-015-3853-3",
    journal = "Eur. Phys. J. C",
    volume = "76",
    number = "1",
    pages = "13",
    year = "2016"
}

@article{Ruijl:2017dtg,
    author = "Ruijl, Ben and Ueda, Takahiro and Vermaseren, Jos",
    title = "{FORM version 4.2}",
    eprint = "1707.06453",
    archivePrefix = "arXiv",
    primaryClass = "hep-ph",
    month = "7",
    year = "2017"
}

@article{Nogueira:1991ex,
    author = "Nogueira, Paulo",
    title = "{Automatic Feynman Graph Generation}",
    reportNumber = "IFM-7-91",
    doi = "10.1006/jcph.1993.1074",
    journal = "J. Comput. Phys.",
    volume = "105",
    pages = "279--289",
    year = "1993"
}

@article{vonManteuffel:2012np,
  author        = {von Manteuffel, A. and Studerus, C.},
  title         = {{Reduze 2 - Distributed Feynman Integral Reduction}},
  eprint        = {1201.4330},
  archiveprefix = {arXiv},
  primaryclass  = {hep-ph},
  reportnumber  = {ZU-TH-01-12, MZ-TH-12-03, BI-TP-2012-02},
  month         = {1},
  year          = {2012}
}

@article{Lange:2025fba,
    author = "Lange, Fabian and Usovitsch, Johann and Wu, Zihao",
    title = "{Kira 3: integral reduction with efficient seeding and optimized equation selection}",
    eprint = "2505.20197",
    archivePrefix = "arXiv",
    primaryClass = "hep-ph",
    reportNumber = "ZU-TH 39/25, HU-EP-25/17-RTG",
    doi = "10.1016/j.cpc.2025.109999",
    journal = "Comput. Phys. Commun.",
    volume = "322",
    pages = "109999",
    year = "2026"
}

@article{Heller:2021qkz,
    author = "Heller, Matthias and von Manteuffel, Andreas",
    title = "{MultivariateApart: Generalized partial fractions}",
    eprint = "2101.08283",
    archivePrefix = "arXiv",
    primaryClass = "cs.SC",
    reportNumber = "MITP/21-002, MSUHEP-20-016",
    doi = "10.1016/j.cpc.2021.108174",
    journal = "Comput. Phys. Commun.",
    volume = "271",
    pages = "108174",
    year = "2022"
}

@article{Kinoshita:1962ur,
  author  = {Kinoshita, T.},
  title   = {{Mass singularities of Feynman amplitudes}},
  doi     = {10.1063/1.1724268},
  journal = {J. Math. Phys.},
  volume  = {3},
  pages   = {650--677},
  year    = {1962}
}

@article{Lee:1964is,
  author  = {Lee, T. D. and Nauenberg, M.},
  editor  = {Feinberg, G.},
  title   = {{Degenerate Systems and Mass Singularities}},
  doi     = {10.1103/PhysRev.133.B1549},
  journal = {Phys. Rev.},
  volume  = {133},
  pages   = {B1549--B1562},
  year    = {1964}
}

@article{Catani_1998,
  title         = {The singular behaviour of QCD amplitudes at two-loop order},
  volume        = {427},
  issn          = {0370-2693},
  url           = {http://dx.doi.org/10.1016/S0370-2693(98)00332-3},
  doi           = {10.1016/s0370-2693(98)00332-3},
  number        = {1–2},
  journal       = {Physics Letters B},
  publisher     = {Elsevier BV},
  author        = {Catani, Stefano},
  collaboration = {},
  year          = {1998},
  month         = may,
  pages         = {161–171}
}

@article{Laporta:2000dsw,
    author = "Laporta, S.",
    title = "{High-precision calculation of multiloop Feynman integrals by difference equations}",
    eprint = "hep-ph/0102033",
    archivePrefix = "arXiv",
    doi = "10.1142/S0217751X00002159",
    journal = "Int. J. Mod. Phys. A",
    volume = "15",
    pages = "5087--5159",
    year = "2000"
}

@article{Gehrmann:1999as,
    author = "Gehrmann, T. and Remiddi, E.",
    title = "{Differential equations for two-loop four-point functions}",
    eprint = "hep-ph/9912329",
    archivePrefix = "arXiv",
    reportNumber = "TTP-99-49",
    doi = "10.1016/S0550-3213(00)00223-6",
    journal = "Nucl. Phys. B",
    volume = "580",
    pages = "485--518",
    year = "2000"
}

@article{Duhr:2019tlz,
    author = "Duhr, Claude and Dulat, Falko",
    title = "{PolyLogTools {\textemdash} polylogs for the masses}",
    eprint = "1904.07279",
    archivePrefix = "arXiv",
    primaryClass = "hep-th",
    reportNumber = "CP3-19-17, CERN-TH-2019-045, SLAC-PUB-17423",
    doi = "10.1007/JHEP08(2019)135",
    journal = "JHEP",
    volume = "08",
    pages = "135",
    year = "2019"
}

@article{Rainwater:1997dg,
    author = "Rainwater, David L. and Zeppenfeld, D.",
    title = "{Searching for $H\to\gamma\gamma$ in weak boson fusion at the LHC}",
    eprint = "hep-ph/9712271",
    archivePrefix = "arXiv",
    reportNumber = "MADPH-97-1023",
    doi = "10.1088/1126-6708/1997/12/005",
    journal = "JHEP",
    volume = "12",
    pages = "005",
    year = "1997"
}

@article{Brein:2012ne,
    author = "Brein, Oliver and Harlander, Robert V. and Zirke, Tom J. E.",
    title = "{vh@nnlo - Higgs Strahlung at hadron colliders}",
    eprint = "1210.5347",
    archivePrefix = "arXiv",
    primaryClass = "hep-ph",
    doi = "10.1016/j.cpc.2012.11.002",
    journal = "Comput. Phys. Commun.",
    volume = "184",
    pages = "998--1003",
    year = "2013"
}

@article{Anastasiou:2002yz,
    author = "Anastasiou, Charalampos and Melnikov, Kirill",
    title = "{Higgs boson production at hadron colliders in NNLO QCD}",
    eprint = "hep-ph/0207004",
    archivePrefix = "arXiv",
    reportNumber = "SLAC-PUB-9273",
    doi = "10.1016/S0550-3213(02)00837-4",
    journal = "Nucl. Phys. B",
    volume = "646",
    pages = "220--256",
    year = "2002"
}

@article{Ravindran:2006bu,
    author = "Ravindran, V. and Smith, J. and van Neerven, W. L.",
    title = "{QCD threshold corrections to di-lepton and Higgs rapidity distributions beyond $N^{2}$ LO}",
    eprint = "hep-ph/0608308",
    archivePrefix = "arXiv",
    reportNumber = "HRI-04-2006, YITP-SB-06-35",
    doi = "10.1016/j.nuclphysb.2007.01.005",
    journal = "Nucl. Phys. B",
    volume = "767",
    pages = "100--129",
    year = "2007"
}

@article{Graudenz:1992pv,
    author = "Graudenz, D. and Spira, M. and Zerwas, P. M.",
    title = "{QCD corrections to Higgs boson production at proton proton colliders}",
    reportNumber = "DESY-92-149, LBL-33154",
    doi = "10.1103/PhysRevLett.70.1372",
    journal = "Phys. Rev. Lett.",
    volume = "70",
    pages = "1372--1375",
    year = "1993"
}

@article{Dawson:1990zj,
    author = "Dawson, S.",
    title = "{Radiative corrections to Higgs boson production}",
    reportNumber = "PRINT-91-0092 (BNL), BNL-45446",
    doi = "10.1016/0550-3213(91)90061-2",
    journal = "Nucl. Phys. B",
    volume = "359",
    pages = "283--300",
    year = "1991"
}

@article{Djouadi:1991tka,
    author = "Djouadi, A. and Spira, M. and Zerwas, P. M.",
    title = "{Production of Higgs bosons in proton colliders: QCD corrections}",
    reportNumber = "PITHA-91-6",
    doi = "10.1016/0370-2693(91)90375-Z",
    journal = "Phys. Lett. B",
    volume = "264",
    pages = "440--446",
    year = "1991"
}

@article{Harlander:2001is,
    author = "Harlander, Robert V. and Kilgore, William B.",
    title = "{Soft and virtual corrections to proton proton ---{\ensuremath{>}} H + x at NNLO}",
    eprint = "hep-ph/0102241",
    archivePrefix = "arXiv",
    reportNumber = "BNL-HET-01-6",
    doi = "10.1103/PhysRevD.64.013015",
    journal = "Phys. Rev. D",
    volume = "64",
    pages = "013015",
    year = "2001"
}

@article{Ravindran:2003um,
    author = "Ravindran, V. and Smith, J. and van Neerven, W. L.",
    title = "{NNLO corrections to the total cross-section for Higgs boson production in hadron hadron collisions}",
    eprint = "hep-ph/0302135",
    archivePrefix = "arXiv",
    reportNumber = "YITP-SB-03-02, INLO-PUB-01-03",
    doi = "10.1016/S0550-3213(03)00457-7",
    journal = "Nucl. Phys. B",
    volume = "665",
    pages = "325--366",
    year = "2003"
}

@article{Anastasiou:2004xq,
    author = "Anastasiou, Charalampos and Melnikov, Kirill and Petriello, Frank",
    title = "{Higgs boson production at hadron colliders: Differential cross sections through next-to-next-to-leading order}",
    eprint = "hep-ph/0409088",
    archivePrefix = "arXiv",
    reportNumber = "SLAC-PUB-10673",
    doi = "10.1103/PhysRevLett.93.262002",
    journal = "Phys. Rev. Lett.",
    volume = "93",
    pages = "262002",
    year = "2004"
}
